\documentclass[a4paper,%
 reprint,
 amsmath,amssymb,
 aps,
 prapplied,
 superscriptaddress,
]{revtex4-2}

\usepackage{graphicx}
\usepackage{dcolumn}
\usepackage{bm}
\usepackage{float}

\begin{document}


\title{Vibrational modes as the origin of dielectric loss at 0.27--100 THz in a-SiC:H
}

\author{B. T. Buijtendorp}
\email[]{b.t.buijtendorp@tudelft.nl}
\affiliation{Faculty of Electrical Engineering, Mathematics and Computer Science, Delft University of Technology, Mekelweg 4, Delft 2628 CD, The Netherlands}

\author{A. Endo}
\affiliation{Faculty of Electrical Engineering, Mathematics and Computer Science, Delft University of Technology, Mekelweg 4, Delft 2628 CD, The Netherlands}

\author{W. Jellema}
\affiliation{SRON Netherlands Institute for Space Research, Landleven 12, Groningen 9747 AD, The Netherlands} 

\author{K. Karatsu}
\affiliation{SRON Netherlands Institute for Space Research, Niels Bohrweg 4, Leiden 2333 CA, The Netherlands}

\author{K. Kouwenhoven}
\affiliation{Faculty of Electrical Engineering, Mathematics and Computer Science, Delft University of Technology, Mekelweg 4, Delft 2628 CD, The Netherlands}
\affiliation{SRON Netherlands Institute for Space Research, Niels Bohrweg 4, Leiden 2333 CA, The Netherlands}

\author{D. Lamers}
\affiliation{SRON Netherlands Institute for Space Research, Niels Bohrweg 4, Leiden 2333 CA, The Netherlands}

\author{A. J. van der Linden}
\affiliation{SRON Netherlands Institute for Space Research, Niels Bohrweg 4, Leiden 2333 CA, The Netherlands}

\author{K. Rostem}
\affiliation{NASA/Goddard Space Flight Center, Code 665, Greenbelt, MD 20771, USA}

\author{H. M. Veen}
\affiliation{SRON Netherlands Institute for Space Research, Niels Bohrweg 4, Leiden 2333 CA, The Netherlands}

\author{E. J. Wollack}
\affiliation{NASA/Goddard Space Flight Center, Code 665, Greenbelt, MD 20771, USA}

\author{J. J. A. Baselmans}
\affiliation{Faculty of Electrical Engineering, Mathematics and Computer Science, Delft University of Technology, Mekelweg 4, Delft 2628 CD, The Netherlands}
\affiliation{SRON Netherlands Institute for Space Research, Niels Bohrweg 4, Leiden 2333 CA, The Netherlands}

\author{S. Vollebregt}
\affiliation{Faculty of Electrical Engineering, Mathematics and Computer Science, Delft University of Technology, Mekelweg 4, Delft 2628 CD, The Netherlands}


\begin{abstract}
Low-loss deposited dielectrics are beneficial for the advancement of superconducting integrated circuits for astronomy. In the microwave band ($\mathrm{\sim}$1--10 GHz) the dielectric loss at cryogenic temperatures and low electric field strengths is dominated by two-level systems. However, the origin of the loss in the millimeter-submillimeter band ($\mathrm{\sim}$0.1--1 THz) is not understood. We measured the loss of hydrogenated amorphous SiC (\mbox{a-SiC:H}) films in the 0.27--100 THz range using superconducting microstrip resonators and Fourier-transform spectroscopy. The agreement between the loss data and a Maxwell-Helmholtz-Drude dispersion model suggests that vibrational modes above 10 THz dominate the loss in the \mbox{a-SiC:H} above 200 GHz.
\end{abstract}

\maketitle

\section{Introduction}
Deposited dielectrics with low loss at millimeter-submillimeter (mm-submm) wavelengths are beneficial for the development of superconducting integrated circuits for astronomy, such as filter banks \cite{taniguchiDESHIMADevelopmentIntegrated2022,pascuallagunaTerahertzBandPassFilters2021, karkareFullArrayNoisePerformance2020, hailey-dunsheathOpticalMeasurementsSuperSpec2014}, on-chip Fourier-transform spectrometers \cite{basuthakurDevelopmentSuperconductingOnchip2023}, and  kinetic inductance parametric amplifiers \cite{tanOperationKineticinductanceTravelling2024}. Although it is possible to fabricate microstrip lines using crystalline Si \mbox{(c-Si)} extracted from a silicon-on-insulator wafer by a flip-bonding process \cite{patelFabricationMKIDSMicroSpec2013}, deposited dielectrics allow for simpler and more flexible chip designs and fabrication routes. Recently there have been multiple reports of excess loss in deposited dielectrics in the mm-submm band ($\mathrm{\sim}$0.1--1 THz) compared to the microwave band ($\mathrm{\sim}$1--10 GHz) \cite{buijtendorpHydrogenatedAmorphousSilicon2022, hahnleSuperconductingMicrostripLosses2021a, oconnellMicrowaveDielectricLoss2008a, endoOnchipFilterBank2013} and it was found that the mm-submm loss increases monotonically with frequency \cite{buijtendorpHydrogenatedAmorphousSilicon2022, endoOnchipFilterBank2013}.  The frequency dependence of the loss is surprising in the framework of the standard tunneling model (STM) for two-level systems (TLSs) \cite{mullerUnderstandingTwolevelsystemsAmorphous2019, phillipsTunnelingStatesAmorphous1972}. Although the STM successfully explains the dielectric loss in the microwave band at cryogenic temperatures and low electric field strengths, the origin of the mm-submm loss is not understood.

We measured the loss of hydrogenated amorphous SiC \mbox{(\mbox{a-SiC:H})} films in the 0.27--100 THz range, where in the 270--600 GHz range we used superconducting microstrip resonators, and above 3 THz we used Fourier-transform spectroscopy (FTS). The agreement between the loss data and an Maxwell-Helmholtz-Drude (MHD) dispersion model suggests that vibrational modes above 10 THz dominate the loss in the \mbox{a-SiC:H} above 200 GHz.

\section{Deposition details and material characterization of the \mbox{\MakeLowercase{a}-S\MakeLowercase{i}C:H}}\label{depo}
\begin{figure}
\includegraphics[width=\linewidth]{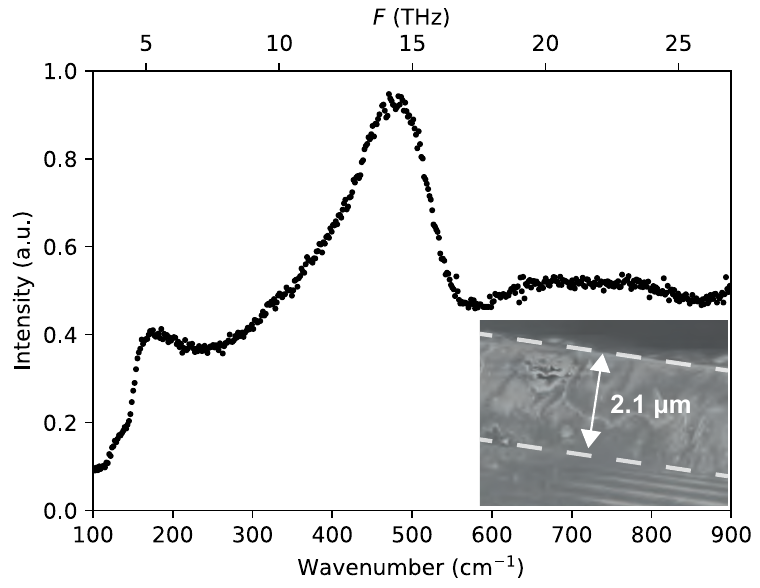}
\caption{\label{fig:raman} Raman spectrum of a 2.1-$\mathrm{\mu m}$-thick \mbox{a-SiC:H} film on a c-Si substrate. The Raman spectrum is typical for fully amorphous \mbox{a-SiC:H}. The inset shows a scanning electron micrograph of a cleaved c-Si wafer with an \mbox{a-SiC:H} film, from which we determined the film thickness.}
\end{figure}
We deposited the 2.1-$\mathrm{\mu m}$-thick \mbox{a-SiC:H} films by plasma-enhanced chemical vapor deposition (PECVD) using a Novellus Concept One \cite{sarroLowstressPECVDSiC1998}. All the \mbox{a-SiC:H} films which we measured for this work were deposited together on a total of four c-Si wafers (W1--4). We fabricated the microstrip resonators on W1, and on this wafer we deposited the \mbox{a-SiC:H} on a NbTiN ground layer that was sputter-deposited on the c-Si substrate. On W2--4 we deposited the \mbox{a-SiC:H} directly on the c-Si substrates. Prior to the deposition of the \mbox{a-SiC:H} we removed any native oxide on each wafer with a 10-s 10\% $\mathrm{HF}$ dip. We deposited the films at a substrate temperature of 400$\mathrm{^{\circ}}$C, $\mathrm{SiH_4}$ flow of 25 sccm, $\mathrm{CH_4}$ flow of 411 sccm, chamber pressure of 2 Torr, 450-kHz RF power of 150 W, and a 13.56-MHz RF power of 450 W. 
    
We performed several characterization measurements on W2 to determine the material properties of the \mbox{a-SiC:H} films. We verified that the \mbox{a-SiC:H} is fully amorphous by performing Raman spectroscopy with a 514-nm laser. We plotted the Raman spectrum in Fig. \ref{fig:raman}. The spectrum's broadened features are typical for fully amorphous \mbox{a-SiC:H} \cite{ricciardiMicrostructureAnalysisASiC2006, yuStructuralOpticalProperties2004, nakashimaRamanInvestigationSiC1997}. We determined that the \mbox{a-SiC:H} has an atomic \mbox{$\mathrm{Si}$ / $\mathrm{C}$} ratio of 0.8 using energy-dispersive X-ray spectroscopy (EDS). We measured a 2.1-$\mathrm{\mu m}$ film thickness from a scanning-electron micrograph of the cross section of the wafer, as shown in the inset of Fig. \ref{fig:raman}. Finally, we measured a relative dielectric constant $\varepsilon_\mathrm{r}^{'}$ of 7.6 at 250 THz and a band gap of 1.8 eV by performing ellipsometry and fitting a Tauc-Lorentz dispersion model \cite{budaiEllipsometricStudySixC2011}.

\section{Far infrared loss, measured in superconducting microstrip resonators}
\subsection{Device design and fabrication}\label{devicedesign}
To measure the 270--600 GHz loss of the \mbox{a-SiC:H} we used a lab-on-chip experiment based upon superconducting Fabry-Pérot (FP) resonators fabricated from \mbox{a-SiC:H} and NbTiN. The design of the loss measurement device (Fig. \ref{fig:chip}a) is similar to the one reported in Refs. \cite{hahnleSuperconductingMicrostripLosses2021a, buijtendorpHydrogenatedAmorphousSilicon2022}, with the exception that we enabled wide-band measurements by using leaky-wave antennas (Fig. \ref{fig:chip}b) fabricated on 1-$\mathrm{\mu m}$-thick $\mathrm{SiN}_x$ membranes \cite{hahnleUltrawidebandLeakyLens2020, netoUWBNonDispersive2010}. The signal from each antenna is coupled to a FP via a NbTiN coplanar waveguide (CPW) with 2-$\mathrm{\mu m}$ center line width and  2-$\mathrm{\mu m}$ gap width, designed to eliminate radiation loss \cite{hahnleSuppressionRadiationLoss2020d}. The four FPs (FP1--4) have a NbTiN/a-SiC:H/NbTiN microstrip geometry with 2.1-$\mathrm{\mu m}$-thick \mbox{a-SiC:H} and 14-$\mathrm{\mu m}$ line width. The power transmitted through the FPs is measured using CPW NbTiN-Al hybrid MKIDs \cite{janssenHighOpticalEfficiency2013}. The couplings between the FPs and the input/output CPW lines are made at the ends of each FP, with a 22-$\mathrm{\mu m}$-long overlap between the FP microstrip lines and the center lines of the NbTiN CPWs (Fig. \ref{fig:chip}d). The galvanic connections between the NbTiN center lines, which are coupled to the FPs and the Al center lines of the CPW NbTiN-Al hybrid MKIDs, are made on a layer of $\mathrm{SiN}_x$ (Fig. \ref{fig:chip}d) to avoid enhanced erosion of the Al at the Si-NbTiN interfaces \cite{ferrariAntennaCoupledMKID2018}.

We optimized the FP resonator and FP coupler design to cover the wide bandwidth of 270--600 GHz. We limited the maximum value of the FP resonators' coupling quality factor $Q_\mathrm{c}$ such that (1) the FP peaks do not become too sharp to measure with the 20 MHz resolution of the photomixer source that generates the 270--600 GHz radiation, and (2) there is sufficient transmission at higher frequencies where the losses are high. Furthermore, we limited the minimum value of $Q_\mathrm{c}$ so that the FP transmission peaks do not have too much overlap to be individually resolved. Taking these limitations into consideration, we chose resonator lengths of 3.00 mm, 4.50 mm, 6.75 mm and 10.13 mm and a coupler overlap length of 22 $\mathrm{\mu m}$ (Fig. \ref{fig:chip}d), resulting in $Q_\mathrm{c}$ values in the range of $0.4\times 10^3$ to $24\times 10^3$ in the range of 600--270 GHz for all four FPs.

We fabricated the device on a high-resistivity c-Si wafer (W1). We coated the wafer with 1 $\mathrm{\mu m}$ of $\mathrm{SiN}_x$ on the front side and on the back side, using low pressure chemical vapor deposition. We removed the $\mathrm{SiN}_x$ front side everywhere except at the antennas and at the Al lines of the MKIDs. We sputter-deposited the 300-nm-thick NbTiN ground layer (under the \mbox{a-SiC:H}), the 100-nm thick NbTiN top layer (on top of the \mbox{a-SiC:H}), and the 100-nm thick Al layer using an Evatec LLS 801. We deposited the \mbox{a-SiC:H} on top of the NbTiN ground layer, as discussed in Section \ref{depo}. We etched the \mbox{a-SiC:H} using reactive ion etching (RIE) with an $\mathrm{SF_6}$ and $\mathrm{O_2}$ plasma. We patterned the leaky-wave antennas, CPW mm-submm feed lines, MKIDs and CPW microwave readout lines into the NbTiN ground plane using RIE. We wet-etched the Al center lines of the MKIDs. We patterned the $\mathrm{SiN}_x$ on the back side using RIE to define KOH etching windows behind the four antennas, and we removed the c-Si substrate below the antennas by KOH etching, such that each antenna sits on top of a $\mathrm{SiN}_x$ membrane. Subsequently, for stray light absorption, we covered the back side of the chip with a $\beta$--Ta mesh \cite{baselmansEliminatingStrayRadiation2018} which we sputter-deposited using an Evatec LLS 801. We removed the $\beta$--Ta below the $\mathrm{SiN_x}$ membranes using RIE.
\subsection{Measurement}
\begin{figure}
\includegraphics[width=\linewidth]{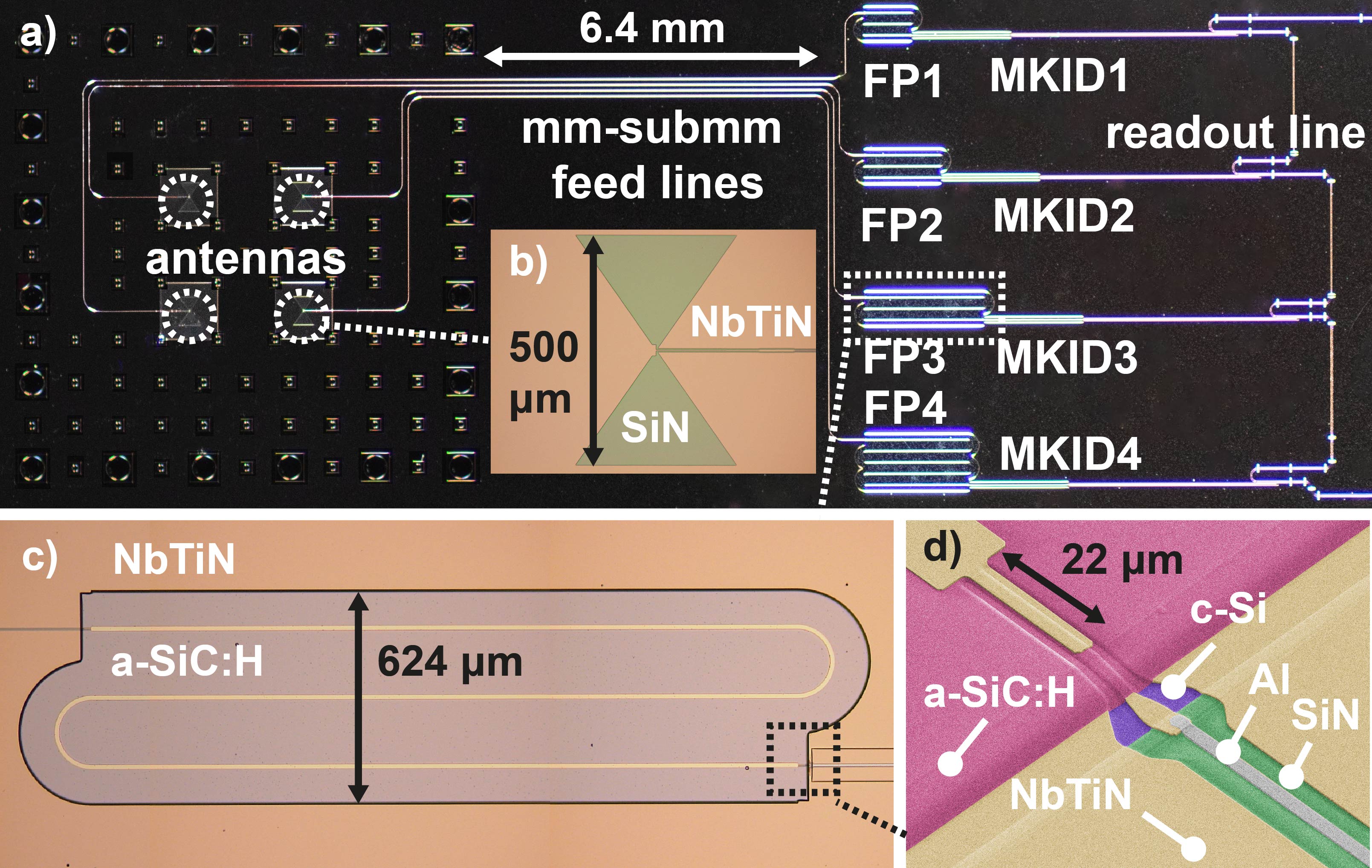}
\caption{\label{fig:chip} a) Photograph of the loss measurement device. The FPs receive the mm-submm signal from antennas, which receive radiation from a Toptica Photonics TeraScan 780 photomixer source. Each FP is coupled to an MKID, which detects the power transmitted through the FP. b) Micrograph of one of the leaky-wave antennas. c) Micrograph of FP3. The feed line is visible on the left side of the image and the MKID is visible on the right side of the image. d)  Tilted scanning electron micrograph of the coupling of FP3 to the MKID, with false coloring.}
\end{figure}
We measured the loss at 20 mK on the chip that is described in Subsection \ref{devicedesign} and shown in Fig. \ref{fig:chip}a. The experimental setup was similar to the one reported in Refs. \cite{hahnleSuperconductingMicrostripLosses2021a, buijtendorpHydrogenatedAmorphousSilicon2022}. The radiation from a continous-wave photomixer source (Toptica Photonics Terascan 780) was swept from 270 to 600 GHz in frequency steps of 20 MHz. The broad beam from the photomixer source was weakly coupled to all four antennas. The loss of the \mbox{a-SiC:H} was determined from the power that was transmitted through the FPs, measured by the response of the MKIDs as a function of frequency. The transmitted power versus frequency is a sum of Lorentzian peaks located at $f_\mathrm{peak} = nf_0$, where $n$ is the mode number and $f_0$ is the fundamental resonance frequency. In Fig. \ref{fig:peaks}a--d we plotted the measured MKID response in a frequency band around 350 GHz. Each peak corresponds to a loaded quality factor $Q=f_\mathrm{peak}/\Delta f$, where $\Delta f$ is the peak's full width at half maximum (FWHM).
\subsection{Results}
\begin{figure}
\includegraphics[width=\linewidth]{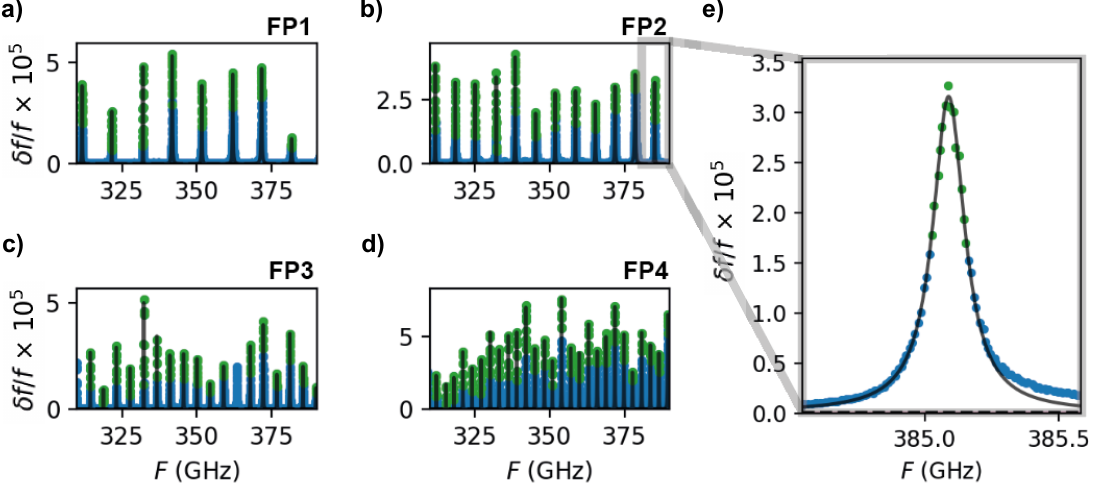}
\caption{\label{fig:peaks} a--d) The measured MKID frequency response of FP1--4 in the 310–380 GHz band. The blue and  green points represent the measurement data. The data represented as green points were used in the fitting. The black curves are Lorentzian fits to the FP transmission peaks, from which we obtained the loaded quality factor $Q$ of each FP peak. The same analysis was performed for each of the seven frequency bands centered around 270, 310, 350, 400, 455, 520 and 600 GHz. e) Close-up of a single FP transmission peak of FP2.}
\end{figure}
\begin{figure}[!ht]
\includegraphics[width=\linewidth]{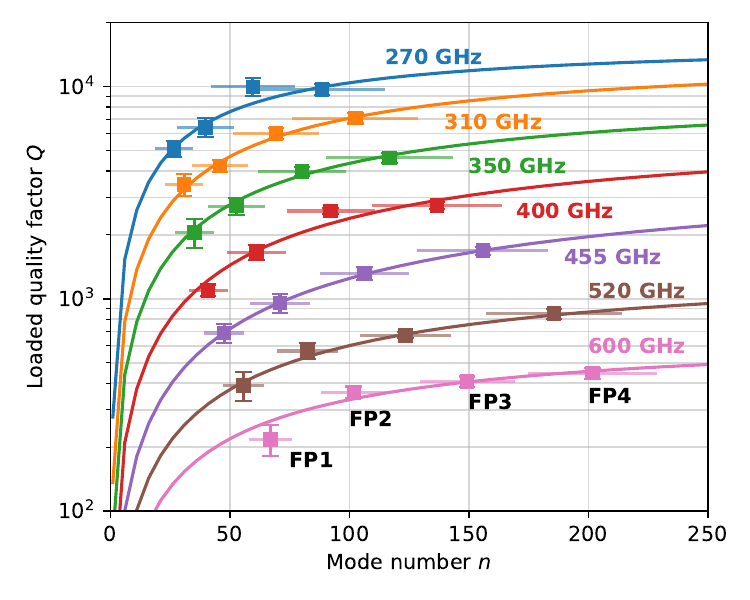}
\caption{\label{fig:Q} Average loaded quality factor $Q$ versus average mode number $n$ of the Fabry-Perot resonators (FP1–4) The curves are fits of Eq. \eqref{eq:Ql_n}. The vertical error bars represent the standard deviation ($\pm \sigma$) in $Q$. The horizontal bars represent the range of mode numbers used in computing the average.}
\end{figure}
We determined the 270--600 GHz loss by fitting Lorentzian peaks to the MKID response data. We plotted the measured MKID response in the 310--380 GHz frequency band in Fig. \ref{fig:peaks}. The variation in the peak heights in Fig. \ref{fig:peaks} is caused by a variation in the mm-submm power that enters the FPs, due to a frequency-dependence of the power output of the photomixer source and due to standing waves between the antennas and the FPs. We determined the $Q$ value of each peak from a Lorentzian fit (Fig. \ref{fig:peaks}e), where the fitting parameter $Q$ is independent from the peak height. For the next steps in the analysis we separated the frequency response data into frequency bands centered around 270, 310, 350, 400, 455, 520 and 600 GHz. We determined the $Q$ and $n$ values at the center frequencies of each band from the average values computed over all peaks within each band. We present the average $Q$ versus average $n$ data in Fig. \ref{fig:Q}. 

\begin{figure}[]
\includegraphics[width=\linewidth]{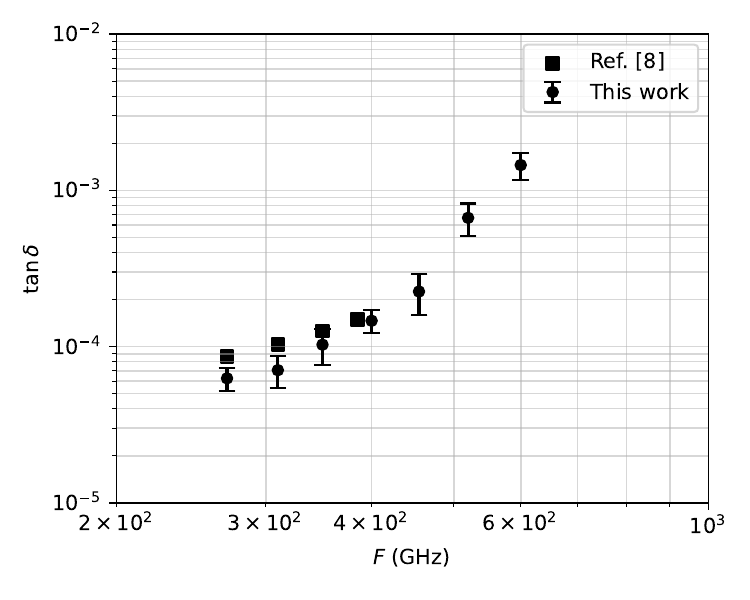}
\caption{\label{fig:mmsubmmloss} The mm-submm $\tan \delta = 1/\left(p Q_\mathrm{i} \right)$ versus frequency, which we obtained by fitting Eq. \eqref{eq:Ql_n} to the data in Fig. \ref{fig:Q}. The points represent the measured $\tan \delta$ values from this work, and the error bars represent one standard deviation ($\pm \sigma$) uncertainty in $\tan \delta$. The squares show the \mbox{a-SiC:H} losses reported in Ref. \cite{buijtendorpHydrogenatedAmorphousSilicon2022}.}
\end{figure}

We determined the $Q_\mathrm{i}$ from the $Q$ versus $n$ data by fitting the equation:
\begin{equation}\label{eq:Ql_n}
    Q = \frac{nQ_\mathrm{c,1}Q_\mathrm{i}}{nQ_\mathrm{c,1} + Q_\mathrm{i}}
,\end{equation}
where $Q_\mathrm{c} = nQ_\mathrm{c,1}$ is the FP's coupling quality factor. Here, the $Q_\mathrm{c,1} = \pi/|t_\mathrm{c}|^2$, where $t_\mathrm{c}$ the transmission coefficient of the FPs' couplers. The FPs have identical couplers and therefore share a single $Q_\mathrm{c,1}$ (and hence $t_\mathrm{c}$) value, which is obtained together with $Q_\mathrm{i}$ from the fit (Fig. \ref{fig:Q}) of Eq. \eqref{eq:Ql_n}. Finally, from the FP resonators' internal quality factor $Q_\mathrm{i}$ we obtained the loss tangent $\tan \delta$:
\begin{equation}\label{eq:tand}
\tan \delta = \left(pQ_\mathrm{i} \right)^{-1}   
,\end{equation}
where $p$ is the filling fraction of the \mbox{a-SiC:H}, which we determined to be 0.97 in our FPs using the EM-field solver Sonnet \cite{SonnetUserGuide}. We present the resulting loss versus frequency data in the 270--600 GHz range in Fig. \ref{fig:mmsubmmloss}. We observe that the loss increases monotonically with frequency and that it is in agreement with the loss reported in Ref. \cite{buijtendorpHydrogenatedAmorphousSilicon2022}.
\section{Mid infrared loss measured by Fourier-transform spectroscopy}\label{sec:FTS}
\subsection{Sample fabrication}\label{sec:samples}
To measure the 3--100 THz loss of the \mbox{a-SiC:H} we fabricated three kinds of \mbox{a-SiC:H} samples for FTS measurements. Sample 1 (S1): A single \mbox{a-SiC:H} membrane  with 2.1 $\mathrm{\mu m}$ thickness and 2 cm $\times$ 2 cm surface area, supported by a 400-$\mathrm{\mu m}$-thick c-Si frame. Sample 2 (S2): A stack of two units that are equivalent to S1, as shown in Fig. \ref{fig:membrane}b. Sample 3 (S3): A 2.1-$\mathrm{\mu m}$-thick \mbox{a-SiC:H} film (not a membrane) on a double side polished c-Si substrate. We fabricated S1 and S2 from W3 and S3 from W4. The \mbox{a-SiC:H} films that were fabricated into membranes were deposited on the front side of the substrates, and were deposited simultaneously with the other \mbox{a-SiC:H} samples that are described in this work. Separately, we deposited an \mbox{a-SiC:H} layer on the back side of W3. This back side layer was removed using RIE below the membrane positions, thereby creating square-shaped etching windows. The membranes were then created by removing the c-Si substrate using KOH etching from the back side of the wafer.

\subsection{Measurement}\label{sec:FTS_method}
\begin{figure}
\includegraphics[width=\linewidth]{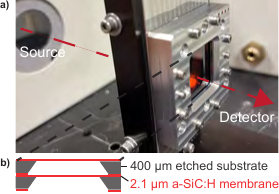}
\caption{\label{fig:membrane} a) Photograph of S2 in the BioRad FTS system. b) Schematic of S2, which consists of two stacked \mbox{a-SiC:H} membranes.}
\end{figure}

To obtain the 3–100 THz loss, we measured the FTS transmisison of S1--3 at room temperature. The FTS transmission is defined as $T\equiv I_\mathrm{s}/I_\mathrm{bg}$. The $I_\mathrm{s}$ is the measured intensity when the sample is in the path from the source to the detector (Fig. \ref{fig:membrane}a), and  $I_\mathrm{bg}$ is the measured intensity without a sample in place. We measured the transmission through S1 and S2 in the ranges of 3--10 THz and 18--30 THz using a BioRad FTS system, at a resolution of 2.0 $\mathrm{cm^{-1}}$ and 0.5 $\mathrm{cm^{-1}}$ for S1 and S2 respectively. In Fig. \ref{fig:membrane}a we present a photograph of S2 inside the BioRad FTS. Here we used a room temperature deuterated triglycine sulfate (DTGS) detector for frequencies above 4 THz and a cryogenic bolometer detector for frequencies below 4 THz. The sample chamber of the BioRad FTS was continously purged with dry air prior to and during the measurement. Furthermore, we measured the transmission through S3 in the range of 18--100 THz using a Thermo Fischer Nicolet FTS system at a resolution of 4.0 $\mathrm{cm^{-1}}$, where the sample chamber was continuously purged with nitrogen prior to and during the measurement.
\subsection{Results}\label{sec:FTS_results}
\begin{figure*}
\includegraphics[width=\textwidth]{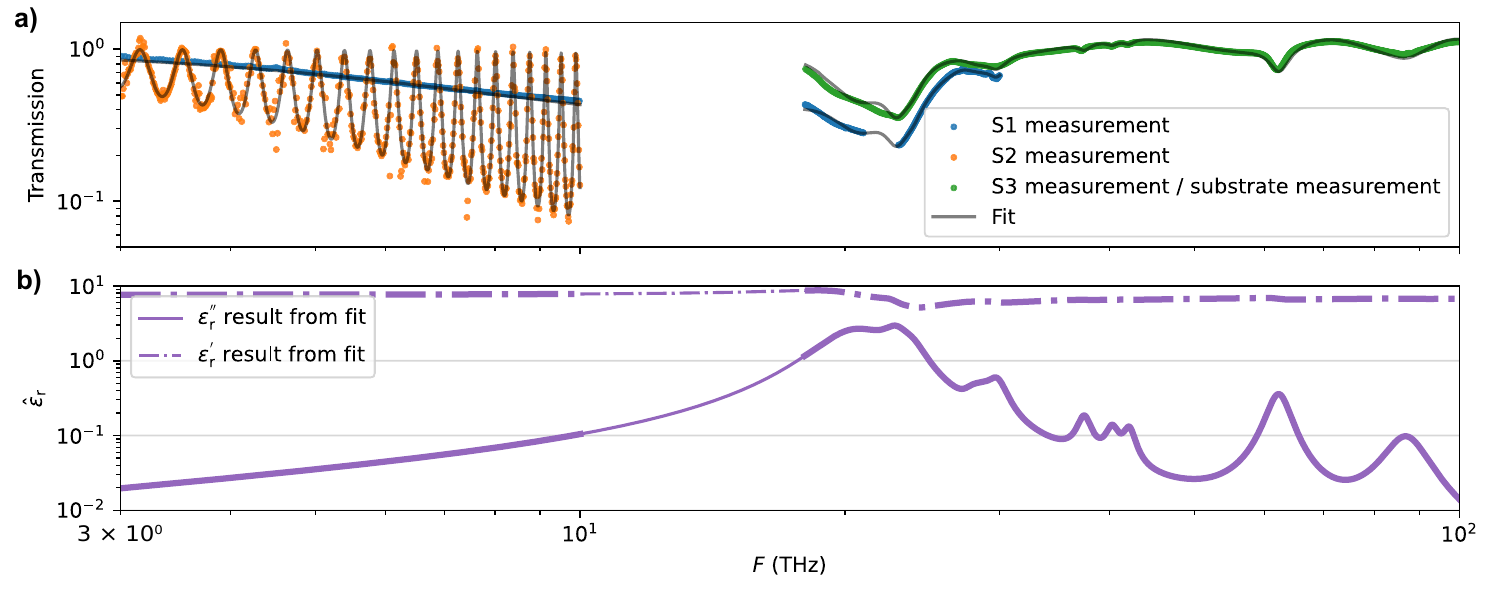}
\caption{\label{fig:FTS} a) The points show the transmission through the \mbox{a-SiC:H} samples measured using FTS in the range of 3--100 THz, for S1--3. The solid grey curves show the best-fit results of TMM models where the $\hat{\varepsilon}_\mathrm{r}$ is parameterized using the MHD dispersion model (Eq. \eqref{eq:maxwell}). The plotted data of S3 was divided by the measured transmission of the bare c-Si substrate to clearly show the absorption features of the \mbox{a-SiC:H}. b) The real ($\varepsilon_\mathrm{r}^{'}$) and imaginary ($\varepsilon_\mathrm{r}^{''}$) parts of the complex permittivity $\hat{\varepsilon}_\mathrm{r}$ resulting from the fitting to the transmission data in Fig. \ref{fig:FTS}a. The thin curves show the fitting results in the range of 3-100 THz, and the thick curves show the ranges where measurement data was available. The annotations identify the modes that are present in the absorption spectrum, which are detailed in Appendix \ref{appA}.}
\end{figure*}

We determined the 3--100 THz loss by fitting transfer matrix method (TMM) \cite{luceTMMFastTransferMatrix2022} models to the FTS transmission data of S1--3. The transmission data and the best-fit transmission curves are presented in Fig. \ref{fig:FTS}a. Here the complex permittivity $\hat{\varepsilon}_\mathrm{r} \equiv \varepsilon_\mathrm{r}^{'} + i\varepsilon_\mathrm{r}^{''}$ (and therefore the loss) of the \mbox{a-SiC:H} was a model parameter which we obtained from the fitting. We combined the FTS transmission data of S1--3 in a single fitting procedure, where we normalized the fitting residuals by the number of data points in each data set. In the case of S3, we measured the bare c-Si substrate prior to the deposition of the \mbox{a-SiC:H} and we included the substrate properties in the TMM model. We note that the periodicity in the transmission of S2 (Fig. \ref{fig:FTS}a ) is dictated by the inter-membrane distance.

In the fitting of the TMM models to the transmission data we parameterized the $\hat{\varepsilon}_\mathrm{r}$ of the \mbox{a-SiC:H} using the MHD dispersion model \cite{cataldoInfraredDielectricProperties2012b, cataldoInfraredDielectricProperties2016b}:
\begin{equation}\label{eq:maxwell}
    \hat{\varepsilon}_\mathrm{r} = \hat{\varepsilon}_\mathrm{\infty} + \sum^{M}_{j=1}\frac{\Delta\hat{\varepsilon}_j\cdot\omega^2_{T_j}}{\omega^2_{T_j}-\omega^2-i\omega\Gamma^{'}_j\left( \omega\right)},
\end{equation}
where the $\hat{\varepsilon}_\mathrm{r}$ consists of a sum of $M$ modes. Here, the broadening $\Gamma^{'}_j$ equals:
\begin{equation}
    \Gamma^{'}_j=\Gamma_j \mathrm{exp}\left[-\alpha_j \left( \frac{\omega^2_{T_j} - \omega^2}{\omega\Gamma_j} \right)^2 \right],
\end{equation}
where the broadening is purely Lorentzian if $\alpha_j$ equals zero, and is an interpolation between Lorentzian and Gaussian if $\alpha_j$ is greater than zero \cite{cataldoInfraredDielectricProperties2012b,kimModelingOpticalDielectric1992}. The $\Gamma_j$ is the Lorentzian damping coefficient, $\omega_{T_j}$ is the center frequency of a mode, and $\Delta\hat{\varepsilon}_j$ is defined as $\Delta\hat{\varepsilon}_j \equiv \hat{\varepsilon}_j - \hat{\varepsilon}_{j+1}$. The modes are ordered such that $\omega_{T_{j+1}} > \omega_{T_{j}}$. The $\hat{\varepsilon}_\infty \equiv \hat{\varepsilon}_{M+1}$ is the contribution from modes at higher frequencies. 

In Fig. \ref{fig:FTS}b we present the $\hat{\varepsilon}_\mathrm{r}$ of the \mbox{a-SiC:H} resulting from the best-fit parameters, which are listed in Table \ref{tab:1} in Appendix \ref{appA}. The observed vibrational modes have previously been reported in literature \cite{kingFourierTransformInfrared2011}: The Si--C stretch absorption band ($\mathrm{\sim}$18--30 THz), the Si--$\mathrm{CH_2}$--Si wagging mode ($\mathrm{\sim}$30--31 THz), the Si--$\mathrm{CH_2}$--Si scissor and bending modes ($\mathrm{\sim}$40 THz) and the C--$\mathrm{H_3}$ symmetric ($\mathrm{\sim}$37 THz) and asymmetric ($\mathrm{\sim}$42 THz) bending bands, the Si--$\mathrm{H}_x$ stretch band ($\mathrm{\sim}$60--69 THz) and the C--$\mathrm{H}_x$ stretch band ($\mathrm{\sim}$84--93 THz). The purple curve of Fig. \ref{fig:totalmodel} represents the $\tan\delta = \varepsilon_\mathrm{r}^{''}/\varepsilon_\mathrm{r}^{'}$ corresponding to the $\hat{\varepsilon}_\mathrm{r}$ (purple curve) plotted in Fig. \ref{fig:FTS}b.
\section{Model of dielectric loss from microwave to mid-infrared}\label{sec:totalmodel}
Finally, we examined whether the loss measured with the FPs at 270--600 GHz and at 20 mK (Fig. \ref{fig:mmsubmmloss}) and the loss measured by FTS in the range of 3--100 THz and at room temperature (fig. \ref{fig:FTS}b) can be consistently explained with a common MHD dispersion model. Here we note that amorphous dielectrics typically exhibit negligible temperature dependence of the dielectric loss at infrared wavelengths \cite{stromTemperatureFrequencyDependences1977}. We searched for a set of parameters in Eq. \eqref{eq:maxwell} using a numerical optimization algorithm that simultaneously minimizes the differences between (1) the calculated $\tan{\delta}$ and the FP-measured $\tan{\delta}$, and (2) the calculated transmission and the FTS-measured transmission. In the fitting we included the FP-measured $\tan{\delta}$ values from Ref. \cite{buijtendorpHydrogenatedAmorphousSilicon2022} (Fig. \ref{fig:mmsubmmloss}). The 520 GHz and 600 GHz points in Fig. \ref{fig:mmsubmmloss} had to be excluded because they could not be reproduced by our current model. The resulting best-fit parameter values are listed in Table \ref{tab:2} in Appendix \ref{appA}. In Fig. \ref{fig:totalmodel} we co-plotted the $\tan \delta$ resulting from the combined fitting to the FTS and FP measurements (grey curve), and the $\tan \delta$ corresponding to Fig. \ref{fig:FTS}b resulting from FTS measurements alone (purple curve). Except for the points at 520 and at 600 GHz (which will be discussed in the next section), the modeled frequency dependence of $\tan \delta$ agrees well with both the FP and FTS measurements over the entire frequency range that was covered. 

\begin{figure*}[ht!]
\includegraphics[width=\textwidth]{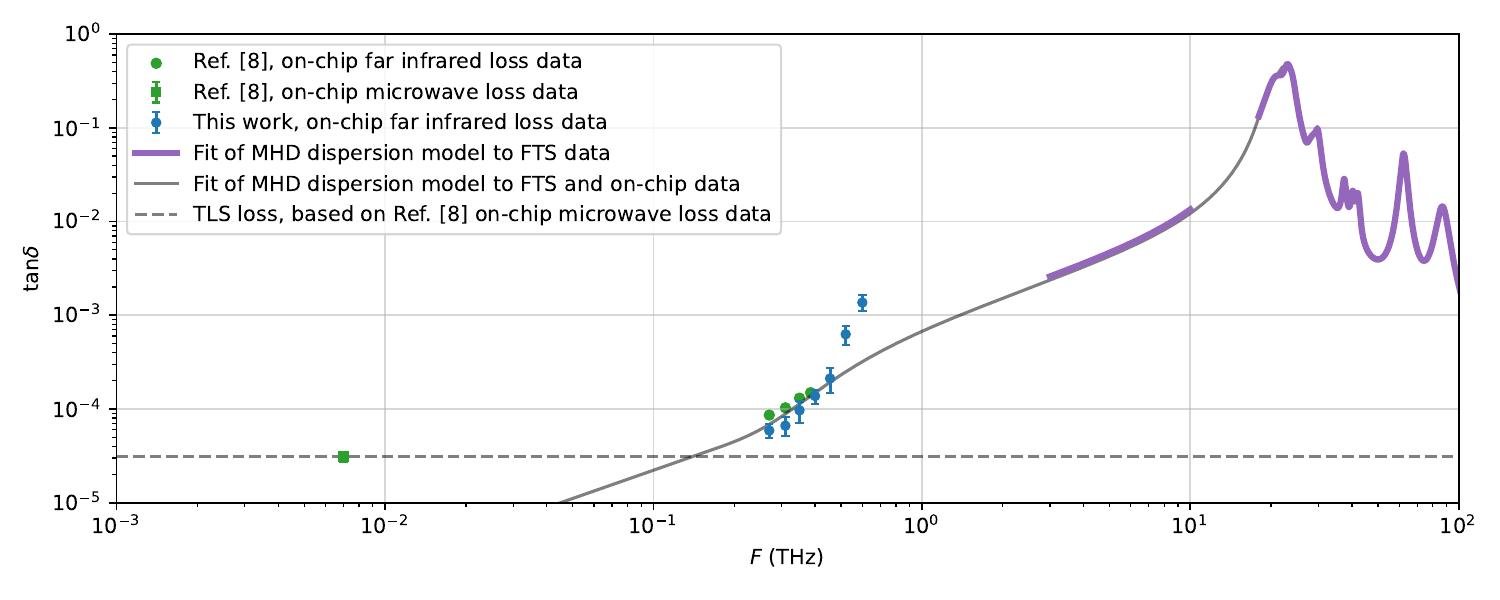}
\caption{\label{fig:totalmodel} Loss tangent $\tan \delta$ versus frequency of the \mbox{a-SiC:H}. The solid grey curve represents the $\tan \delta$ which we obtained by fitting the MHD dispersion model to the combined FTS-measured transmission data (Fig. \ref{fig:FTS}a) and FP-measured loss data (Fig. \ref{fig:mmsubmmloss}), as described in Section \ref{sec:totalmodel}. The purple curve represents the $\tan \delta$ which we obtained by fitting the MHD dispersion model (Eq. \eqref{eq:maxwell}) to only the FTS data (Fig. \ref{fig:FTS}a). The blue points represent the FP-measured loss data from this work (Fig. \ref{fig:mmsubmmloss}). The green points represent the \mbox{a-SiC:H }loss data reported in Ref. \cite{buijtendorpHydrogenatedAmorphousSilicon2022}. The intersection of the solid grey curve and the horizontal dashed line illustrates the crossover between the low-power cryogenic TLS loss and the loss due to vibrational modes.}
\end{figure*}

\section{Discussion on the origin of the loss in deposited dielectrics}
Our results show that the absorption tail of vibrational modes located at frequencies above 10 THz can explain the measured losses in the range of 270--455 GHz. Above a frequency of roughly 200 GHz the losses are anticipated to be dominated by vibrational modes, and below a frequency of roughly 200 GHz and at cryogenic temperatures and low power the loss is expected to be dominated by TLSs. At 520 GHz and at 600 GHz the on-chip loss data is not reproducible by the MHD dispersion model with the modes that have been reported in literature. For an overview of the modes in amorphous and crystalline SiC we refer to Refs. \cite{kingFourierTransformInfrared2011, hofmeisterOPTICALCONSTANTSSILICON2009}. The excess loss at 520 GHz and at 600 GHz is not expected to originate from the bulk of the superconductor since the NbTiN has an energy gap of roughly 1.1 THz. However, it is possible that the excess loss is caused by a small fraction of the superconductor which behaves differently than the bulk of the superconductor \cite{lacquanitiThicknessDependenceElectrical1995, minhajThicknessDependenceSuperconducting1994, chandrasekaranNbTextureEvolution2021}. More experiments are needed to clarify this discrepancy between the model and our data at 520 GHz and at 600 GHz.

Our results pave the way for a thorough understanding of the dielectric loss in deposited dielectrics from microwave to near-infrared wavelengths. We emphasize that strong vibrational modes in the 10--30 THz range are present not only in \mbox{a-SiC:H}, but have also been reported for other commonly used deposited dielectrics such as a-Si:H \cite{langfordInfraredAbsorptionStrength1992, buijtendorpCharacterizationLowlossHydrogenated2022a}, $\mathrm{SiN}_x$ \cite{cataldoInfraredDielectricProperties2012b} and $\mathrm{SiO}_x$ \cite{cataldoInfraredDielectricProperties2016b}. Furthermore, the TLS loss tangent values at cryogenic temperatures and low electric field strengths  $\tan\delta_\mathrm{TLS}$ reported at microwave frequencies of a-Si:H ($\mathrm{\sim} 10^{-5}$) \cite{oconnellMicrowaveDielectricLoss2008a, mazinThinFilmDielectric2010, hahnleSuperconductingMicrostripLosses2021a}, $\mathrm{SiN}_x$ ($\mathrm{\sim} 10^{-4}$) \cite{oconnellMicrowaveDielectricLoss2008a} and $\mathrm{SiO}_x$ ($\mathrm{\sim} 10^{-4}$--$10^{-3}$) \cite{oconnellMicrowaveDielectricLoss2008a, gaoMeasurementLossSuperconducting2009b} lead us to anticipate a similar crossover for these materials from loss dominated by TLSs to loss dominated by vibrational modes. 

\section{Conclusion}
We demonstrated that the dielectric loss of \mbox{a-SiC:H} above 200 GHz is dominated by the absorption tail of vibrational modes which are located at frequencies above 10 THz. Our results pave the way for a thorough understanding of the dielectric loss in deposited dielectrics from microwave to near infrared wavelengths.

\begin{acknowledgements}
We thank the staff of the Else Kooi Laboratory and the Kavli Nanolab Delft for their support. This work was supported by the European Union (ERC Consolidator Grant No. 101043486 TIFUUN). Views and opinions expressed are however those of the authors only and do not necessarily reflect those of the European Union or the European Research Council Executive Agency. Neither the European Union nor the granting authority can be held responsible for them.
\end{acknowledgements}
\appendix 
\section{FTS fitting parameters}\label{appA}
\begin{table*}[!htbp]
    \centering
    \begin{tabular}{c|ccccc}
        \hline
         $j$ & Vibrational mode & $\omega_{T,j}/2\pi$ (THz) & $\Delta \varepsilon_j^{'}$ & $\Gamma_j$(THz) & $\alpha_j$ \\
          \hline
         1 & Si--C stretch & 20.7 & 5.5 $\times 10^{-1}$ & 4.6 & $\mathbf{0}$ \\
         2 & Si--C stretch & 22.8 & 9.8 $\times 10^{-2}$ & 1.7 & 3.8 $\times 10^{-2}$ \\
         3 & Si--C stretch & 23.9 & 5.8 $\times 10^{-2}$ & 1.9 & $\mathbf{0}$ \\
         4 & Si--$\mathrm{CH_2}$--Si wag & 28.3 & 9.0 $\times 10^{-3}$ & 1.7 & 4.2 $\times 10^{-1}$  \\
         5 & Si--$\mathrm{CH_2}$--Si wag & 29.8 & 2.5 $\times 10^{-2}$ & 1.8 & $\mathbf{0}$\\
         6 & C--$\mathrm{H_3}$ symmetric bend & 37.4 & $\mathbf{5.0 \times 10^{-3}}$ & $\mathbf{1.5}$ & $2.6 \times 10^{-6}$  \\
         7 & Si--$\mathrm{CH_2}$--Si scissor or symmetric bend & 40.3 & $\mathbf{3.0 \times 10^{-3}}$ & $\mathbf{1.5}$ & $\mathbf{0}$  \\
         8 & C--$\mathrm{H_3}$ asymmetric bend & 42.1 & $\mathbf{3.0 \times 10^{-3}}$ & $\mathbf{1.5}$ &  $\mathbf{0}$  \\
         9 & Si--$\mathrm{H}_x$ stretch & 62.3 & 2.2 $\times 10^{-2}$ & $\mathbf{4.0}$ & $\mathbf{0}$  \\
         10 & C--$\mathrm{H}_x$ stretch & $\mathbf{87.0}$ &1.1 $\times 10^{-2}$ & $\mathbf{10.0}$ & $\mathbf{0}$ \\
    \end{tabular}
    \caption{Best-fit values corresponding to the results in Fig. \ref{fig:FTS}b, of the MHD dispersion model paramaters (Eq. \ref{eq:maxwell}). The MHD dispersion model was fitted to only the FTS-measured transmission data (Fig. \ref{fig:FTS}a). The $\varepsilon_\infty$ resulting from the fit equals 6.8. The values shown in bold were constrained to the listed values to reduce the number of free fitting parameters.}
    \label{tab:1}
\end{table*}
\begin{table*}[h!]
\centering
    \begin{tabular}{c|ccccc}
        \hline
         $j$ & Vibrational mode & $\omega_{T,j}/2\pi$ (THz) & $\Delta \varepsilon_j^{'}$ & $\Gamma_j$(THz) & $\alpha_j$ \\
          \hline
         1 & Si--C stretch & 19.6 & 1.9 $\times 10^{-1}$ & 3.1 & $\mathbf{0}$ \\
         2 & Si--C stretch & 22.1 & 4.8 $\times 10^{-1}$ & 4.2 & $1.5 \times 10^{-5}$ \\
         3 & Si--C stretch & 23.5 & 3.3 $\times 10^{-2}$& 1.2 & $\mathbf{0}$ \\
         4 & Si--$\mathrm{CH_2}$--Si wag & 28.5 & 9.3 $\times 10^{-3}$ & 1.5 & $5.2 \times 10^{-1}$ \\
         5 & Si--$\mathrm{CH_2}$--Si wag & 29.8 & 1.6 $\times 10^{-2}$ & 1.2 & $\mathbf{0}$ \\
         6 & C--$\mathrm{H_3}$ symmetric bend & 37.4 & $\mathbf{5.0 \times 10^{-3}}$& $\mathbf{1.5}$ & $6.2 \times 10^{-7}$ \\
         7 & Si--$\mathrm{CH_2}$--Si scissor or symmetric bend & 40.2 & $\mathbf{3.0 \times 10^{-3}}$ & $\mathbf{1.5}$ & $\mathbf{0}$ \\
         8 & C--$\mathrm{H_3}$ asymmetric bend & 42.1 & $\mathbf{3.0 \times 10^{-3}}$ & $\mathbf{1.5}$ & $\mathbf{0}$ \\
         9 & Si--$\mathrm{H}_x$ stretch & 62.3 & 2.3 $\times 10^{-2}$ & $\mathbf{4.0}$ & $\mathbf{0}$ \\
         10 & C--$\mathrm{H}_x$ stretch & $\mathbf{87.0}$ & 1.1 $\times 10^{-2}$ & $\mathbf{10.0}$ & $\mathbf{0}$ \\
    \end{tabular}
    \caption{Best-fit values corresponding to the solid grey curve in Fig. \ref{fig:totalmodel}, of the MHD dispersion model parameters (Eq. \ref{eq:maxwell}). The MHD dispersion model was fitted to the combined FTS-measured transmission data (Fig. \ref{fig:FTS}a) and the FP-measured loss data (Fig. \ref{fig:mmsubmmloss}). The $\varepsilon_\infty$ resulting from the fit equals 6.8. The values shown in bold were constrained to the listed values to reduce the number of free fitting parameters.}
    \label{tab:2}
\end{table*}
\clearpage
\bibliography{paper}

\end{document}